\documentclass[%
 aip,
 amsmath,amssymb,
 reprint,%
]{revtex4-1}

\usepackage{graphicx}
\usepackage{dcolumn}
\usepackage{bm}
\usepackage[utf8]{inputenc}
\usepackage[T1]{fontenc}
\usepackage{mathptmx}
\usepackage{etoolbox}
\usepackage{comment}
\usepackage[linesnumbered,ruled,vlined]{algorithm2e}

\makeatletter
\def\@email#1#2{%
 \endgroup
 \patchcmd{\titleblock@produce}
  {\frontmatter@RRAPformat}
  {\frontmatter@RRAPformat{\produce@RRAP{*#1\href{mailto:#2}{#2}}}\frontmatter@RRAPformat}
  {}{}
}%
\makeatother
\begin{document}

\preprint{AIP/123-QED}

\title[Neural refractive index field: Unlocking the Potential of Background-oriented Schlieren Tomography in Volumetric Flow Visualization]{Neural refractive index field:\\Unlocking the Potential of Background-oriented Schlieren Tomography in Volumetric Flow Visualization}
\author{Yuanzhe He}
 \affiliation{%
 School of Mechanical Engineering, Shanghai Jiao Tong University, Shanghai 200240, China
 }
\author{Yutao Zheng}%
 \affiliation{%
 School of Mechanical Engineering, Shanghai Jiao Tong University, Shanghai 200240, China
 }%
\author{Shijie Xu}
 \affiliation{%
 School of Mechanical Engineering, Shanghai Jiao Tong University, Shanghai 200240, China
 }%
\author{Chang Liu}
 \affiliation{%
 School of Engineering, University of Edinburgh, Edinburgh EH9 3JL, United Kingdom
}%
\author{Di Peng}
 \affiliation{%
 School of Mechanical Engineering, Shanghai Jiao Tong University, Shanghai 200240, China
 }%
\author{Yingzheng Liu}
 \affiliation{%
 School of Mechanical Engineering, Shanghai Jiao Tong University, Shanghai 200240, China
 }%

\author{Weiwei Cai}
 \email{cweiwei@sjtu.edu.cn.}
 \homepage{https://me.sjtu.edu.cn/teacher_directory1/caiweiwei.html.}
 \affiliation{%
 School of Mechanical Engineering, Shanghai Jiao Tong University, Shanghai 200240, China
 }%

\date{\today}

\begin{abstract}
Background-oriented Schlieren tomography (BOST) is a prevalent method for visualizing intricate turbulent flows, valued for its ease of implementation and capacity to capture three-dimensional distributions of a multitude of flow parameters. However, the voxel-based meshing scheme leads to significant challenges, such as inadequate spatial resolution, substantial discretization errors, poor noise immunity, and excessive computational costs. This work presents an innovative reconstruction approach termed neural refractive index field (NeRIF) which implicitly represents the flow field with a neural network, which is trained with tailored strategies. Both numerical simulations and experimental demonstrations on turbulent Bunsen flames suggest that our approach can significantly improve the reconstruction accuracy and spatial resolution while concurrently reducing computational expenses. Although showcased in the context of background-oriented schlieren tomography here, the key idea embedded in the NeRIF can be readily adapted to various other tomographic modalities including tomographic absorption spectroscopy and tomographic particle imaging velocimetry, broadening its potential impact across different domains of flow visualization and analysis.

\end{abstract}

\maketitle

\begin{quotation}
Neural refractive index field: Unlocking the Potential of Background-oriented Schlieren Tomography in Volumetric Flow Visualization
\end{quotation}


\section{Introduction}
\label{sec:headings}
Flow visualization techniques play a pivotal role in the field of fluid mechanics and combustion dynamics, offering detailed experimental data that is vital for understanding complex flow phenomena, enhancing control strategies, and verifying models in computational fluid dynamics (CFD)~\citep{settles2017review}. The advent of digital cameras and lasers has given rise to a multitude of visualization methods, each with its unique strengths: Particle Imaging Velocimetry (PIV) is renowned for its ability to create two-dimensional velocity maps~\citep{lavoie2007spatial}, providing a snapshot of the flow velocity at a given plane; Planar Laser-Induced Fluorescence (PLIF) is adept at visualizing the distribution of minor chemical components in flames~\citep{tanahashi2005simultaneous}, such as CH and OH, which are important intermediate species in combustion processes; and Filtered Rayleigh Scattering is used for temperature measurements \citep{carpenter2020filtered}, offering insights into the thermal dynamics within the flow, to list a few. 

While these methods are significantly powerful, they often require expensive excitation lasers, complex optical systems, and specialized expertise for implementation. In contrast, Background-Oriented Schlieren (BOS) has gained widespread popularity since its introduction in 2000 due to its ease of implementation and versatility in measuring multiple physical parameters such as density, temperature and refractive index~\citep{dalziel2000whole,raffel2000applicability}. It has been successfully applied in various challenging scenarios to visualize turbulent flows. For example, the solar disk has been used as a background for visualizing flow around a supersonic T-38 aircraft, captured by a ground-based imaging system equipped with a Calcium-K line filter~\citep{hill2017ground}. Similarly, BOS has been used to capture detailed schlieren images of a full-scale supersonic aircraft in flight, using the desert flora as a background~\citep{heineck2021background}, resulting in the most detailed airborne schlieren imagery to date. Recently, the group of Grauer and Cai pioneered the integration of BOS with physics-informed neural networks (PINNs) and successfully demonstrated the method on axis-symmetric turbulent flows~\citep{molnar2024reconstructing,molnar2023physics,molnar2023estimating,caiFlowEspressoCup2021,cai2021physics}. This approach further enhances the capability of BOS as it enforces the measured quantity to be consistent with the governing equations of the flows and allows for the inference of additional physical properties such as velocity and pressure.  

Despite the great success of these BOS methods, they are inherently two-dimensional and struggle to capture the full three-dimensional nature of turbulent flows~\citep{grauer2023volumetric}. Many teams have tried to solve the problem of three-dimensional flow field diagnosis, such as computed chemiluminescence tomography (CTC)~\citep{cai2013three,liu2017demonstration} for  
measurement of chemical components, and tomographic
particle image velocimetry (TPIV) for measurement of the three-dimensional velocity field~\citep{agarwal2017cylinder}.  Recently, Volumetric background-oriented Schlieren tomography (VBOST) has emerged as a promising technique~\citep{goldhahnBackgroundOrientedSchlieren2007,raffelBackgroundorientedSchlierenBOS2015}, which is widely recognized for its capability to retrieve the three-dimensional distribution of a spectrum of physical parameters~\citep{liuVolumetricImagingFlame2021,liuTimeresolvedThreedimensionalImaging2020}. In recent years, significant progress has been made in VBOST, particularly in addressing measurement uncertainties arising from image displacement computations~\citep{atcheson2009evaluation,schmidt2021wavelet}. These uncertainties are often attributed to the first-order Taylor approximation in optical flow algorithms and the discretization process in cross-correlation algorithms~\citep{vinnichenko2023performance}. Efforts to mitigate errors associated with optical flow have also been notable, with advancements in unified VBOST methods~\citep{grauerFastRobustVolumetric2020} and hybrid approaches that combine voxel-based coarse reconstruction with light deflection considerations for fine reconstruction~\citep{huReconstructionRefinementHybrid2024}.

However, the existing VBOST modalities typically rely on a voxel-based meshing scheme, which assumes constant physical properties within each voxel. This approach introduces several limitations, including significant discretization errors that result in severe reconstruction artifacts~\citep{boas2012ct,wang2019high}, exponential scaling of the weight matrix and memory requirements with increased voxel counts, and spatial resolution bottlenecked by a limited number of discrete voxels. To overcome these limitations, A lot of teams have made great efforts. Bo et al. employed a recurrent neural network based on Gated Recurrent Units (GRUs) to mitigate reconstruction artifacts generated during iterative processes to a certain extent.~\citep{boBackgroundorientedSchlierenTomography2023}; Meanwhile, Li et al. utilized the visual hull (VH) method to decrease the number of voxels requiring reconstruction, effectively reducing artifacts while enhancing processing speed~\citep{li2024three}. This work introduces a reconstruction approach termed Neural Refractive Index Field (NeRIF), which aims to train a light-weight neural network in the reconstruction process that receives the inputs of spatial coordinates of the points within the reconstruction domain and outputs both the corresponding refractive index and its gradient. The name of NeRIF draws inspiration from the cutting-edge computer vision algorithm called neural radiance field (NeRF) which has been straightforwardly adapted to tomographic techniques that reconstruct light radiance field ~\citep{mildenhall2021nerf, zhang2023voxel, kelly2023fluidnerf}. Despite the conceptual parallel in utilizing neural networks, the physical models underpinning NeRF and NeRIF are inherently different. While the former focuses on modeling a radiance field, the latter is tailored to handle fields of refractive index and its gradient—highlighting a crucial divergence that requires a unique implementation strategy, marking a pivotal contribution of this research.

To guarantee the superiority of NeRIF, three specially designed mechanisms are implemented. First, the reconstruction domain is implicitly represented with a compact neural network to avoid voxel-based discretization. Second, a tailored random sampling strategy is adopted to approximate the continuous integration and differentiation which are two essential steps for estimating the ray deflection, effectively mitigating discretization error as well as improving the spatial resolution. Third, both the output refractive index and the corresponding gradient are used to estimate the ray deflection for cross-validation, which prevents the neural networks from over-fitting the measurement noise. This step also involves the combination of the automatic and numerical differentiation mechanisms that helps expedite its convergence, thus improving the efficiency of reconstruction~\citep{chiu2022can}. 

Here, the efficacy of NeRIF is rigorously assessed with both numerical studies using synthetic data of turbulent flames from CFD simulations and experimental demonstration on an unpiloted turbulent Bunsen flame. The results indicate that NeRIF can substantially enhance reconstruction accuracy and spatial resolution while concurrently reducing computational costs. 
 
\section {Mathematical formulation of BOST}
\subsection{Voxel-based formulation}
The principle of BOST has been discussed extensively in various literature and only essentials are summarized here for the sake of brevity. As illustrated in Fig.~\ref{fig:schematic} (a), background-oriented schlieren relies on the phenomenon that rays deviate when they propagate through a non-uniform medium with a varied refractive index, causing distortions in the captured background pattern, which contains densely distributed markers. The deflection angle of a ray can be described by
\begin{equation}
\varepsilon = \frac{1}{n}\int_{\text{path}} \nabla n \, ds \approx \sum_{} \Delta s\nabla n ,
\label{eq:integral relation}
\end{equation}
where \emph{n} is the refractive index and \(s\) the integration variable corresponding to the propagation path. The refractive index is related to density according to the Gladstone-Dale equation as
\begin{equation}
n = 1 + \rho K_{G-D}(\lambda),
\label{eq:GD}
\end{equation}
where \(\rho\) represents the density and \(K_{G-D}\) is the Gladstone-Dale constant related to the gas composition.

Experimentally, the LHS of Eq.~\ref{eq:integral relation} can be calculated according to
\begin{equation}
\begin{bmatrix}
\mathbf{\varepsilon_u} \\
\mathbf{\varepsilon_v}
\end{bmatrix}
=
\left( \frac{L_d}{L_b} + 1 \right) 
\frac{1}{f}
\begin{bmatrix}
\Delta \mathbf{u} \\
\Delta \mathbf{v}
\end{bmatrix},
\label{eq:paraxial}
\end{equation}
where \(\Delta \mathbf{u}\) and \(\Delta \mathbf{v}\) are two orthogonal displacement vectors for all rays that can be accurately calculated based on the distorted and the un-distorted images using optical flow algorithms~\citep{grauerInstantaneous3DFlame2018}, \(L_b\) and \(L_d\) are distances between the optical components as shown in Fig.~\ref{fig:schematic} (a), and \(f\) is camera focus.

To solve for the refractive index, the RHS of equation~\ref{eq:integral relation} should be discretized as
\begin{equation}
\begin{bmatrix}
\mathbf{\varepsilon_u}\\
\mathbf{\varepsilon_v}
\end{bmatrix}
=
\mathbf{P}
\begin{bmatrix}
\mathbf{SD_x} \\
\mathbf{SD_y} \\
\mathbf{SD_z}
\end{bmatrix}
 \mathbf{n},
\label{linear equation systems}
\end{equation}
where \(\mathbf{S}\) and \(\mathbf{D}\) represent the pre-calculated tomographic projection matrix and the differential matrix, respectively; and \(\mathbf{P}\) represents the projection matrix that maps the deflection angle from the world coordinate system to the camera coordinate system \citep{grauerInstantaneous3DFlame2018}. 
As we can see, the voxel-based formulation leads to a large system of linear equations, which is typically ill-posed.



\subsection{Neural refractive index field}
The specific implementation process of the algorithm is shown in Algorithm.~\ref{algorithm:1}. The algorithm here employs an implicit neural network-based approach \citep{mildenhall2021nerf} to represent the 3D refractive index field spatially. Specifically, the target function of the refractive index and its corresponding gradient is formulated as \([\mathbf{n_p}, \mathbf{\nabla n_p}] =f(\Theta,\mathbf{p}) \), where \( \mathbf{p}=(x, y, z )\) are input spatial coordinates, and \( \Theta \) represents parameters of the neural network to be trained. Here, both the output refractive index and its gradient from this function are used separately to compute the corresponding sensor plane displacements based on the calibrated camera parameters. The calibration process is well-documented in \citep{grauer2023volumetric}.

\begin{figure}
  \centering
  \includegraphics[width=1\linewidth]{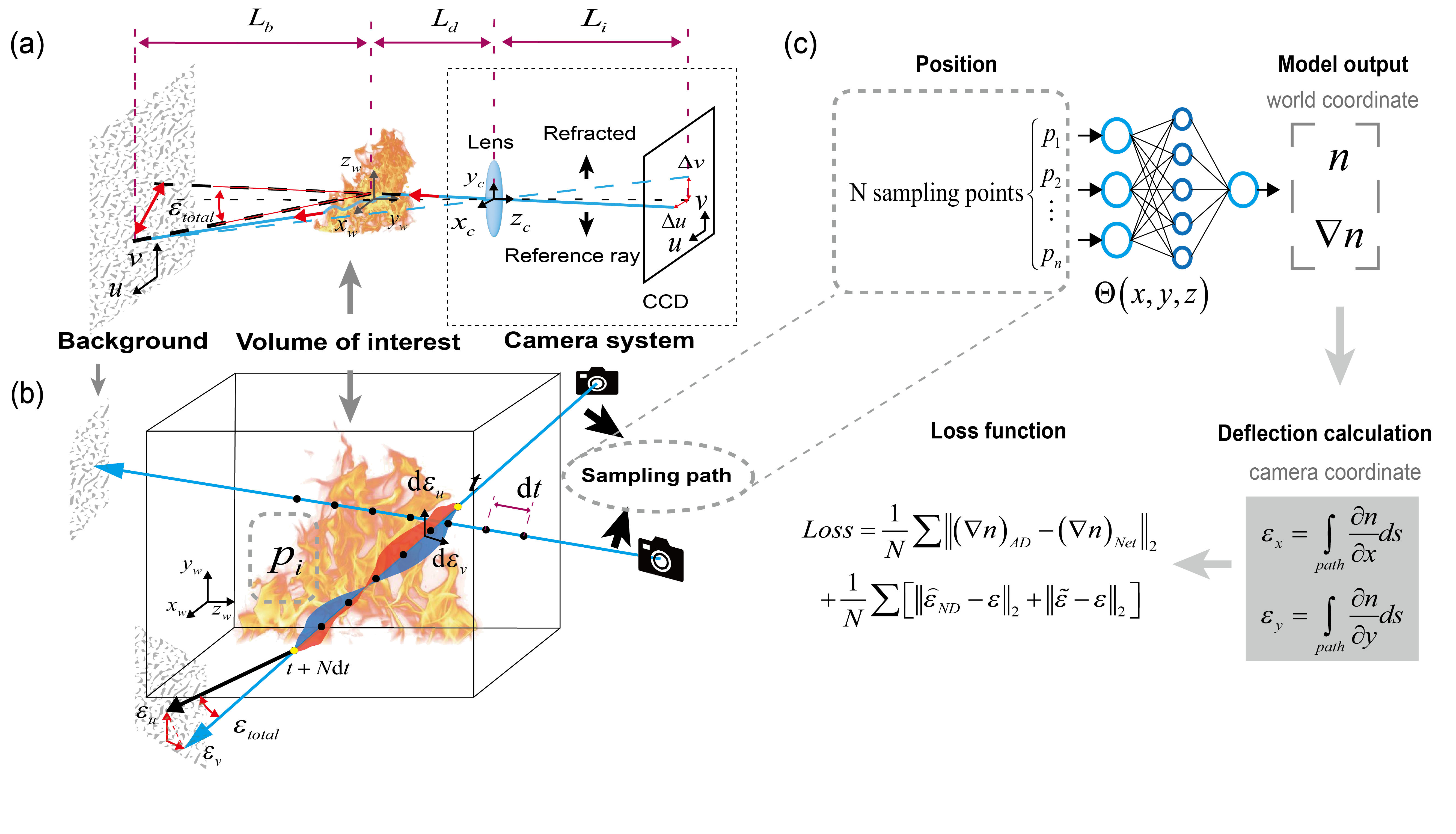}  
  \caption{(a) Principle of background-oriented schlieren; (b) Illustration of integrating the gradient of refractive index along light rays in the proposed algorithm; (c) The reconstruction process of NeRIF.}
  \label{fig:schematic}
\end{figure}

We back-trace the rays from the camera pixel to the background board, as shown in Fig.~\ref{fig:schematic}(b). We can obtain a sequence of refractive indices and gradients by inputting a series of sampling points along the ray path into the neural network. These values are then fed into Eq.~\ref{eq:integral relation} to calculate the displacements.
The integration path theoretically extends from the camera to the background board. To minimize the computational cost, we define the integral range as \([t_n, t_f]\), where \(t_n\) and \(t_f\) represent the distances of the farthest and the nearest ray points from the camera within the reconstruction domain. A random sampling strategy is then applied as follows: First, we set a random sampling number \(N\), which follows a discrete uniform distribution \(N \sim \text{Uniform}(M-10, M+10)\), where \(M\) increases with the number of iterations, ranging from 60 to 200. Then, we divide the integration path into \(N\) sampling intervals and randomly sample a point within each interval as 
\begin{equation}
t_i \sim U\left( t_n + \frac{i-1}{N}(t_f - t_n), t_n + \frac{i}{N}(t_f - t_n) \right), \quad i = 1, 2, \ldots, N
\label{eq:sample}
\end{equation}
where \(t_i\) is the distance between the sampled point \(\mathbf{p_i}\) (see Fig.\ref{fig:schematic} (b)) and the camera pixel, and \(U\) represents continuous uniform sampling. Compared with the voxel-based Monte Carlo sampling algorithm, the sampling interval on the backpropagation light corresponding to each pixel only contributes to the pixel itself, and random sampling ensures accuracy while greatly reducing the calculation amount.

Finally, we apply numerical differentiation (ND) using the central difference method, with the step size \(\Delta\) being selected as
\begin{equation}
\left(\nabla n_{p_i}\right)_{\text{ND}} = \frac{n_{{\mathbf{p}_i}+\Delta \mathbf{r}} - n_{{\mathbf{p}_i}-\Delta \mathbf{r}}}{2 \Delta}, \quad \mathbf{r} = \mathbf{i}, \mathbf{j}, \mathbf{k}, \quad \Delta = \frac{t_f - t_n}{2N}
\label{eq:samplegrad}
\end{equation}
where \(\mathbf{r}\) represents unit vectors in the \emph{x}, \emph{y} and \emph{z} directions. This random sampling strategy helps avoid discretization errors and ensures that the discrete sampling statistically approximates continuous integration and differentiation, reducing the risk of falling into local optima. The specific reconstruction process is illustrated in Fig.~\ref{fig:schematic}(c). 



The constructed loss function is given in Eq.~\ref{eq:lossfunction}. It is crucial to acknowledge that when the coordinates of a sampled point are input into the mult-head neural network, it simultaneously and decoupledly generates both the refractive index and its gradients. However, a key aspect often overlooked is that these outputs are not inherently guaranteed to be consistent with each other. Specifically, if one were to take the derivative of the refractive index \((\nabla n)_{AD}\) output by the network via automatic differentiation (AD) and compare it with the directly predicted refractive index gradient \((\nabla n)_\text{Net}\), the two might not align perfectly. We incorporate their difference into the loss function as the first term to address the mentioned inconsistency. In addition, both the refractive index and its gradients can be independently leveraged to predict displacements. Ideally, the predictions based on both quantities should converge on the same displacement values, matching those measured images experimentally. To enforce this, we introduce two additional terms into the loss function. The first term measures the difference between the integral displacements \(\hat{\varepsilon}_\text{ND}\) predicted from the refractive index using the discrete ND method and the experimentally measured image displacements \({\varepsilon}\). Similarly, the second term quantifies the discrepancy between the integral displacements \(\tilde{\varepsilon}\) calculated from the refractive index gradients output of NeRIF and the same measured data. This approach integrates the mechanisms of ND and AD~\citep{chiu2022can}, enhancing both the robustness and efficiency of the model, while ensuring an accurate representation of the underlying physics.


\begin{equation}
\text{Loss} = \frac{1}{N} \sum \left[  \left\| \left(\nabla n\right)_{\text{AD}} - \left(\nabla n\right)_\text{Net} \right\|_2 + \left\| \hat{\varepsilon}_{ND} - \varepsilon  \right\|_2 + \left\| \tilde{\varepsilon} - \varepsilon \right\|_2 \right].
\label{eq:lossfunction}
\end{equation}

A neural network architecture with 7 fully connected layers of 256 neurons was adopted, incorporating a skip connection at the 4th layer, directly linking it to the input encoded with sinusoidal frequency encoding. The 7th layer connects to four individual single-layer neural networks, each with 64 outputs, to model \(n\) and \(\nabla n\) using a multi-head approach. The Adam optimizer was employed with a learning rate of \(10^{-3}\) and decay rates of (0.9, 0.99) \citep{diederik2014adam},  and more detailed information is provided in Appendix.~\ref{app:NNstructure}.

\begin{algorithm}[H]
\caption{\label{algorithm:1}NeRIF}
\DontPrintSemicolon
Create a multi-head neural network and initialize\;
The displacements \({\varepsilon}\) of images from different views are obtained by optical flow algorithm\;
Trace rays in reverse from image coordinate points \((u,v)\) to obtain sampling points \(p=(x_w,y_w,z_w)\) in the world coordinate system as Eq.~\ref{eq:sample} \;
\While{\( \text{epoch} < n_{\text{outer}} \) and \(\text{loss}>0.005\text{loss}_\text{initial} \)}{
    input p to neural network and output \(n\) and \((\nabla n)_\text{Net}\) \;
    calculate the first term \(\left\| \left(\nabla n\right)_{\text{AD}} - \left(\nabla n\right)_\text{Net} \right\|_2\) of the loss function through AD of \(n\) \;
    calculate \(\hat{\varepsilon}_\text{ND}\) and  \(\tilde{\varepsilon}\) through Eq.~\ref{eq:integral relation} and Eq.~\ref{eq:samplegrad} \;
    backward the total loss as Eq.~\ref{eq:lossfunction} \;
    epoch \( \leftarrow \) epoch + 1\;
}
The final density \(\rho\) is derived from the G-D constant as Eq.~\ref{eq:GD}\;
\end{algorithm}

\section{Numerical and experimental validations}
In the numerical validation process, a spatially randomly distributed 3D Gaussian sphere, designated as Phantom 1, was chosen as the reconstruction target. This selection facilitated a straightforward comparison with the actual refractive index gradient distribution, owing to its differentiable nature. Additionally, temporally resolved spray combustion flames, with a precise time interval of 0.1 ms, spanning from Phantoms 2 to 7 \citep{hadadpour2022effects}, were employed as another validation target. These flames were chosen for their intricate density distribution, particularly characterized by a significant density drop in the product side, which further enhanced the validation's rigor.
\begin{table}
\centering
\caption{Numerical validation settings.}
\begin{tabular}{ll}
\hline
\textbf{Simulated camera system} & \\
Pixel size & 20 $\mu$m $\times$ 20 $\mu$m  \\
Pixel resolution & 200 $\times$ 200 \\
Focal length & 50 mm \\
Number of views & 9 \\
Angle \(\theta\) & \(0^\circ-170^\circ\)\\
\hline
\textbf{Other settings} & \\
Volume of interest & 50 mm $\times$ 50 mm $\times$ 50 mm\\
Noise type & Gaussian\\
\hline
\end{tabular}
\label{tab:parameter}
\end{table}

The fourth-order Runge-Kutta method was utilized for refractive index fields to perform accurate ray tracing, resulting in the generation of nine simulated projection maps (each 200$\times$200 pixels) at equidistant angular intervals spanning from 0 to 170 degrees, details of the implementation of numerical calculations are in Appendix.~\ref{app:RK}. To mimic practical situations, varying levels of Gaussian noise were deliberately introduced into the simulated distorted maps. For comparative studies, the Tikhonov \citep{grauerInstantaneous3DFlame2018} and Landweber \citep{liuVolumetricImagingFlame2021} algorithms were employed without noise, each configured for a 40$\times$40$\times$40 voxel division. The simulation experiments were conducted on a high-performance server equipped with 64 cores of Intel\textsuperscript{\textregistered} Xeon\textsuperscript{\textregistered} Gold CPUs, 512 GB of RAM, and an NVIDIA RTX 4090 GPU, ensuring efficient and reliable computations. Detailed information on the simulation study is shown in Table.~\ref{tab:parameter}.

\begin{figure}
  \centering
  \includegraphics[width=1\linewidth]{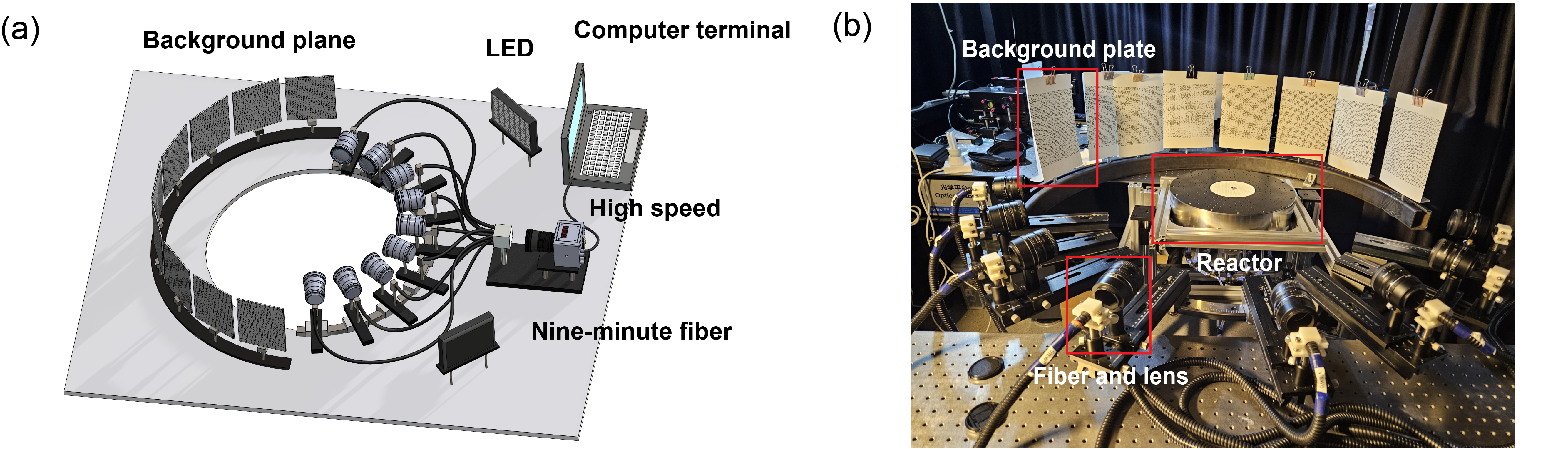}  
  \caption{(a) Schematic diagram of the experimental setup; (b) Reactor and experimental setup objects.}
  \label{fig:setup}
\end{figure}

The experimental implementation of the NeRIF is carried out on a Bunsen flame reactor with a recording system, which consists of a Photron AX100 mini high-speed camera, a fiber bundle with nine input ends, and one output end. Each input end is equipped with a physical focal length of 50 mm lens to register the object image. The background plate and the lens about the shooting area are equally spaced from the layout, as shown in Fig.~\ref{fig:setup}.

The experiments were conducted using a high-speed camera to capture the native flame with a temporal resolution of 1 kHz, which is fast enough to capture the evolution of the flame subjected to perturbation. The Calibration of internal and external parameters for high-speed cameras is carried out by Matlab Camera Calibration Toolbox~\citep{MATLABToolbox}. The optical flow algorithm Deepflow ~\citep{weinzaepfelDeepFlowLargeDisplacement2013} was then selected to infer the displacement vectors according to the comparison between algorithms\citep{grauerFastRobustVolumetric2020}, etc. Detailed information on the experimental study is shown in Table.~\ref{tab:expparameter} and Table.~\ref{tab:flowrate}.
\begin{table}
\centering
\caption{Specifications of the experimental system.}
\begin{tabular}{ll}
\hline
\textbf{Camera} & \\
Exposure time & 1 ms \\
Frame rate & 1 kHz \\
Pixel resolution & 1024 $\times$ 1024 \\
Pixel size & 20 $\mu$m $\times$ 20 $\mu$m \\
Camera Lens & 50 mm and f/2.8 \\
Objective lenses & 24 mm and f/2.8 \\
\hline
\textbf{Fiber bundles} & \\
Resolution of fibers of input ends & 355 $\times$ 420 \\
Dimensionality of the input ends & 5.2 mm $\times$ 5.2 mm \\
Dimensionality of the output end & 16 mm $\times$ 16 mm \\
\hline
\end{tabular}
\label{tab:expparameter}
\end{table}

\begin{table}
 \centering 
 \caption{Flow rates of various gases}
 \begin{tabular}{l|ccc} 
   \hline\hline
   Gas & CH$_4$ & O$_2$ & N$_2$ \\
    
   Flow Rate (SLPM) & 1.00 & 7.93 & 2.00 \\
   \hline\hline
 \end{tabular}
 \label{tab:flowrate}
\end{table}

\section{Results}

\subsection{Numerical validation}
\begin{figure}
  \centering
  \includegraphics[width=1\linewidth]{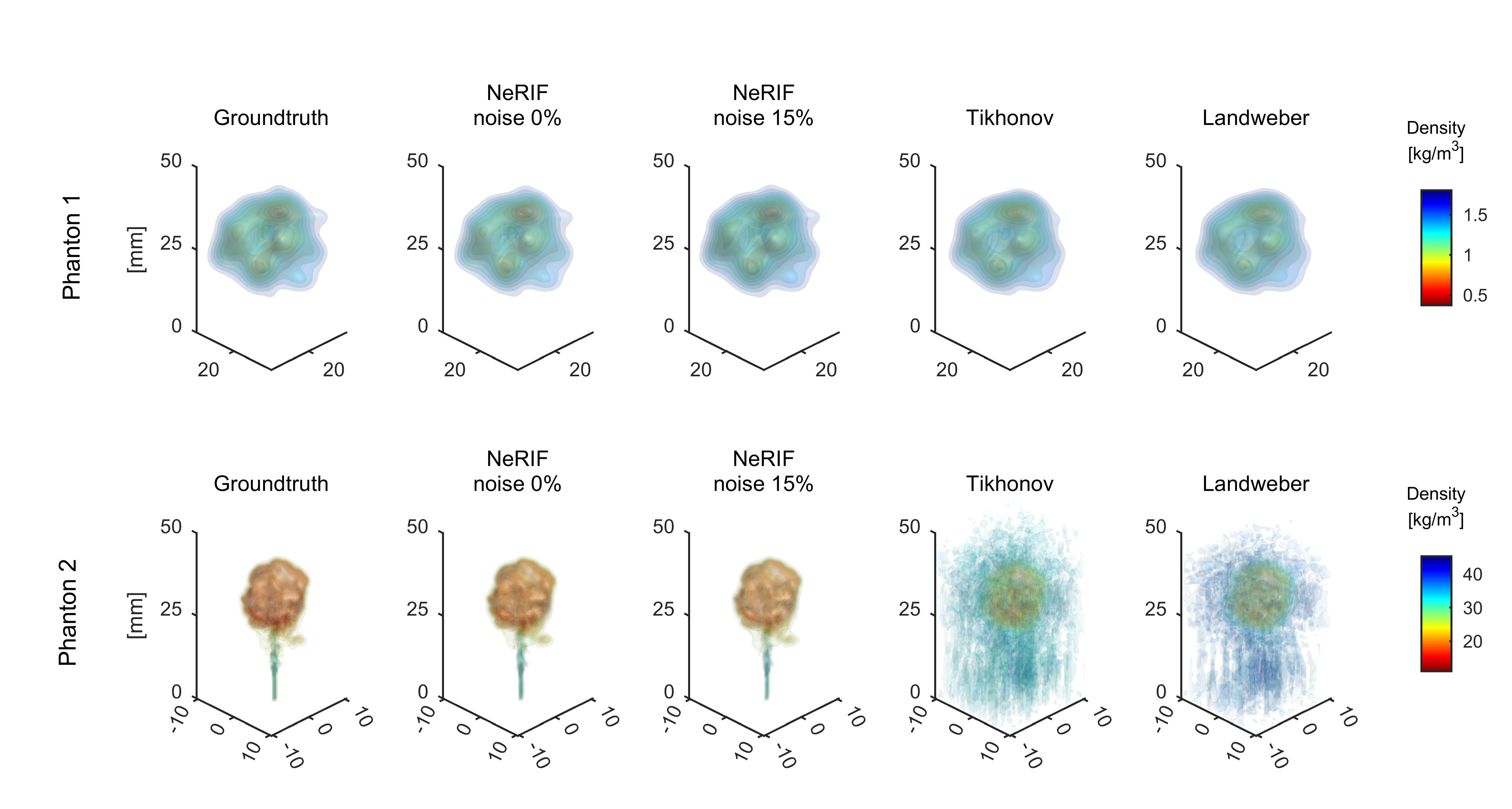}  
  \caption{Reconstruction of 3D density field distribution with different noise levels and algorithms}
  \label{fig:3Dslice}
\end{figure}

\begin{figure}
  \centering
  \includegraphics[width=1\linewidth]{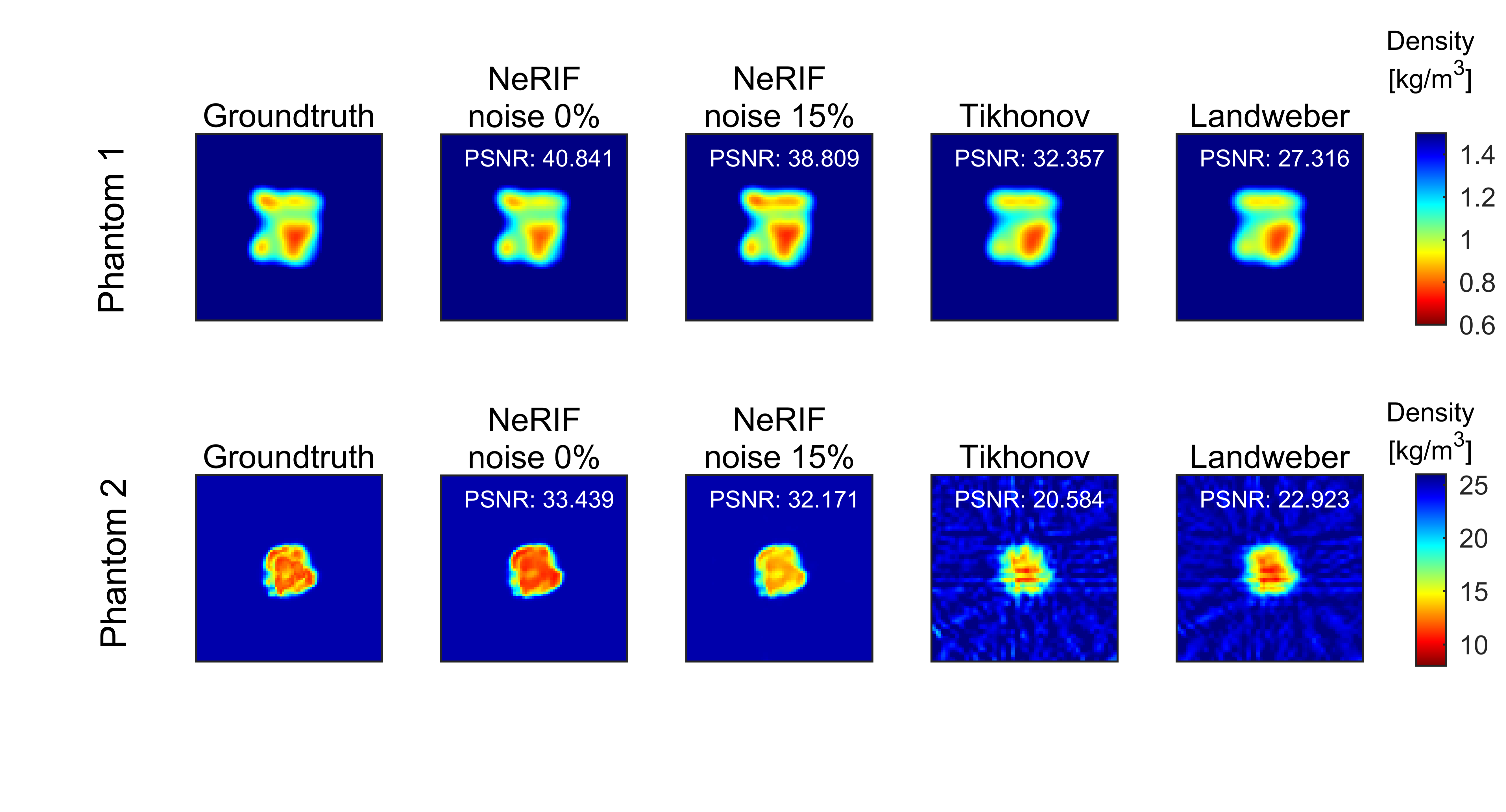}  
  \caption{2D slices of reconstructed density field distribution with different noise levels and algorithms}
  \label{fig:2Dslice}
\end{figure}

Fig.~\ref{fig:3Dslice} showcases the performance of NeRIF under various noise conditions in comparison to the Tikhonov and Landweber algorithms. For Phantom 1 featured with smooth density distribution, it proves challenging to discern differences in the reconstructions. However, in the case of Phantom 2, characterized by larger spatial density gradients and a more intricate, stochastic distribution, NeRIF outperforms the others by reconstructing a 3D density map with minimal visible errors. Conversely, the alternative algorithms exhibit notable reconstruction artifacts, highlighting the superiority of NeRIF in complex scenarios.In the slice graph Fig.~\ref{fig:2Dslice} at z=40mm, we can intuitively see the advantages of NeRIF in reducing reconstruction artifacts.

We undertake a further comparison of the algorithms against seven simulated spray flame scenarios. For each phantom, we analyze and evaluate the reconstructions of ten uniformly spaced slices, spanning a z-range from 5 to 75 mm, utilizing the Structural Similarity Index (SSIM) and correlation coefficient (CC) as assessment metrics, and detailed information is provided in Appendix.~\ref{app:Metrics}. As depicted in Fig.~\ref{fig:2DsliceSSIMCorr}, the voxel-based algorithm exhibits larger errors, with a CC hovering around 0.9, suggesting a moderate level of correlation with the phantoms. Furthermore, the SSIM values generally fall below 0.9, likely attributed to the inherent limitations of voxel discretization, which poses challenges in accurately reconstructing intricate details. In contrast, NeRIF achieves both SSIM and CC values that are approaching 1, underscoring its superior reconstruction quality. Notably, even as noise levels increase, the errors of NeRIF remain lower than those observed with voxel-based algorithms, demonstrating its robustness.

\begin{figure}
  \centering
  \includegraphics[width=\linewidth]{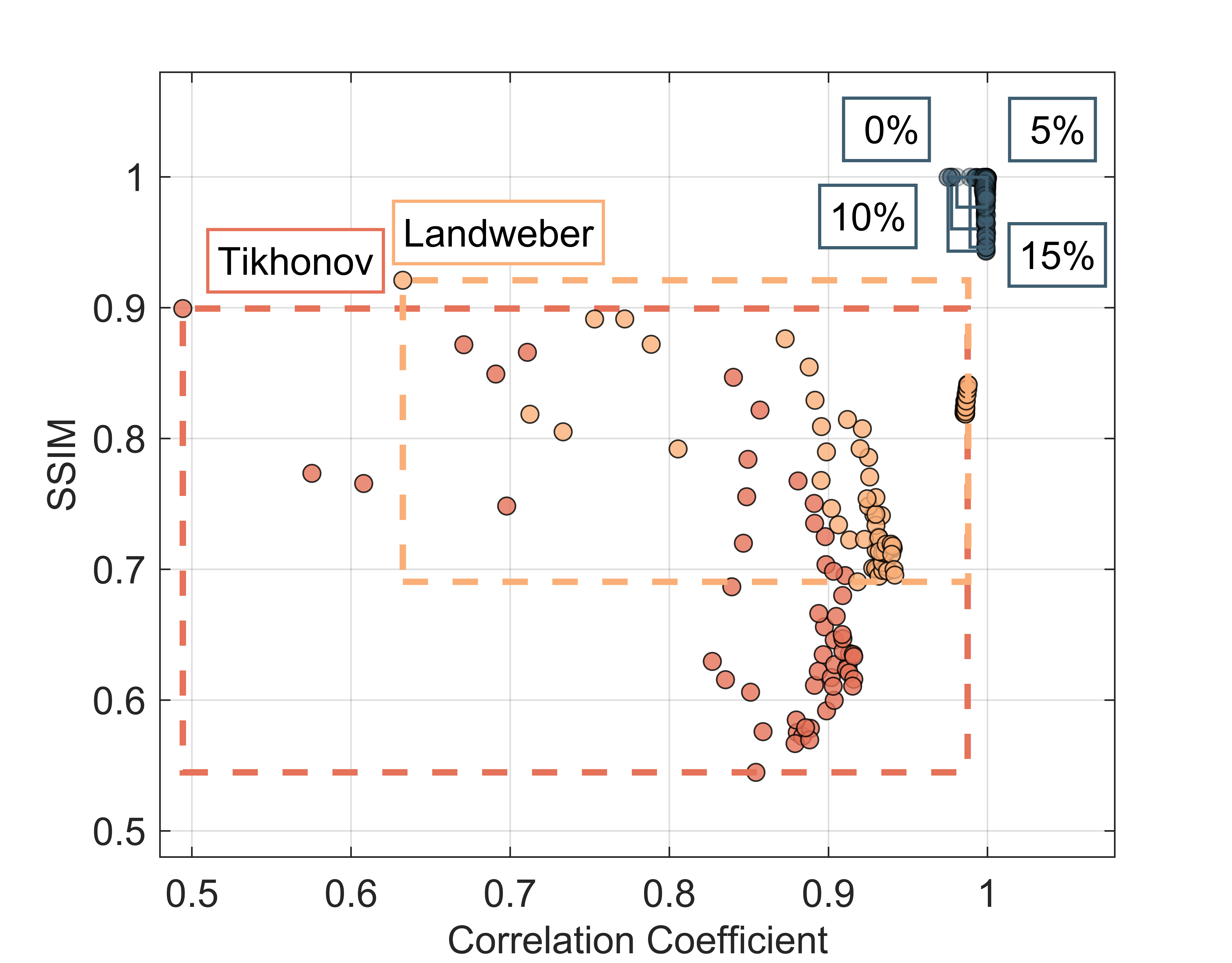}  
  \caption{Structural similarity and correlation coefficients for different slices of each phantom}
  \label{fig:2DsliceSSIMCorr}
\end{figure}
To further quantify the reconstruction quality, the average reconstruction error is defined as \(
\left\| \mathbf{n}_{\text{gt}} - \mathbf{n}_{\text{rec}} \right\|_2 / \left\| \mathbf{n}_{\text{gt}} - 1 \right\|_2
\), termed the L2 error. Table.~\ref{tab:corr_coeffs} and Table.~\ref{tab:l2_errors} compares the algorithms across the entire reconstruction volume under varying noise levels. This comparison unequivocally demonstrates the superiority of NeRIF over voxel-based algorithms in terms of its robustness to noise and overall reconstruction quality. Notably, the CC for the reconstructed density fields using NeRIF consistently exceeds 0.99 across different noise levels, with L2 errors predominantly below 1\%. We quantified the spatial resolution of the algorithm in Appendix.~\ref{app:spatialreslution}.

\begin{figure}
  \centering
  \includegraphics[width=1\linewidth]{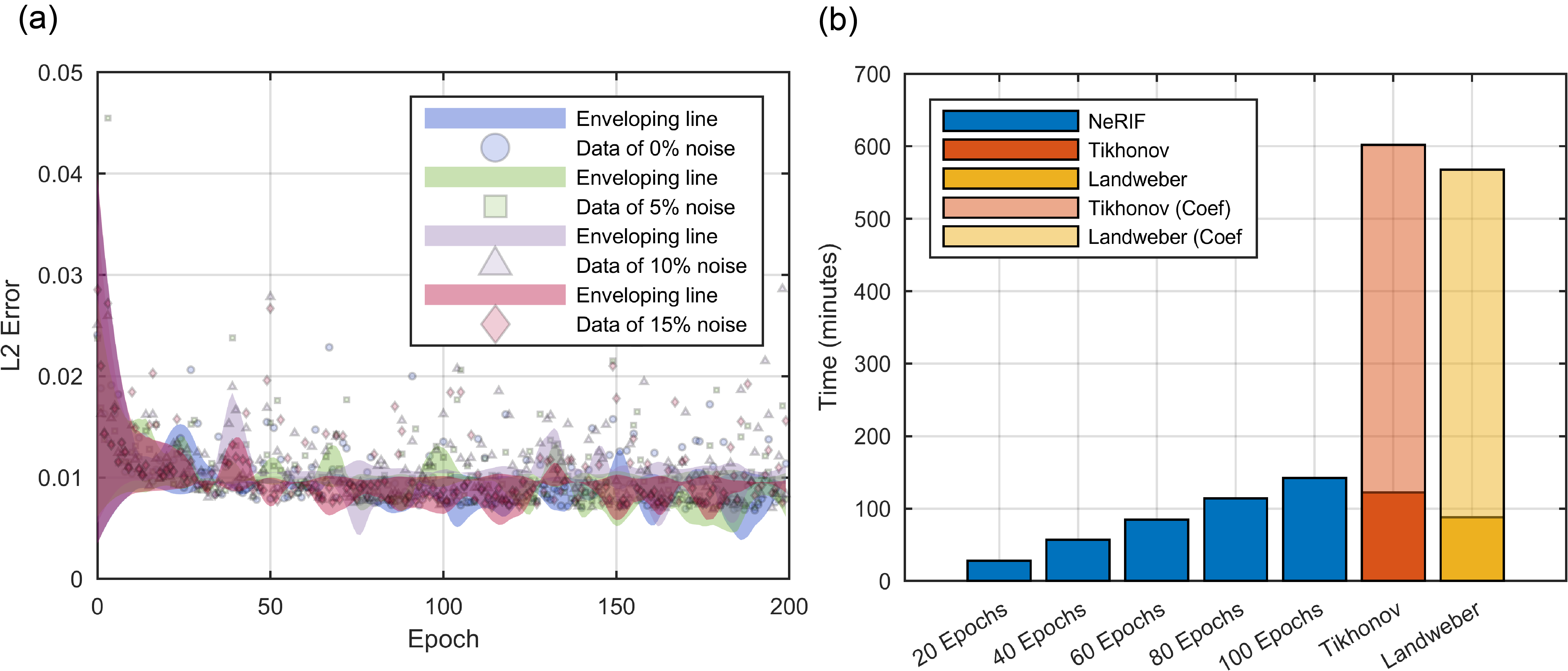}  
  \caption{(a) Errors of the algorithms proposed in this paper under L2 paradigm with different numbers of iterations; (b) Comparison of the temporal running time of different algorithms}
  \label{fig:error_time}
\end{figure}

We delve deeper into the robustness of NeRIF by examining its performance under varying noise levels and iteration numbers, using Phantom 2 as a representative case. As illustrated in Fig.~\ref{fig:error_time}, NeRIF exhibits a rapid reduction in reconstruction error during the initial iterations, with each case converging to an acceptable level of error within approximately 50 epochs. Notably, while the influence of noise becomes more pronounced as iterations progress, the error oscillates slightly but continues to decrease, allowing NeRIF to converge towards a stable residual faster.

On the other hand, the voxel-based reconstruction method faces significant limitations due to the prohibitively large size of the coefficient matrix calculated using the Monte Carlo method \citep{yu2017quantification}, necessitating a reduction in voxel resolution to 40$\times$40$\times$40. Even with this compromise, the computational demands remain immense, according to float32 storage, requiring over 184 GB of storage for the coefficient matrix alone, and consuming over 400 GB of running memory during intermediate computations. These demands not only extend the computation time to over an hour for each projection angle but also render higher voxel resolutions unfeasible due to the cubic increase in computational complexity, which can lead to server memory overflow or excessive runtime.

In contrast, NeRIF not only surpasses voxel-based methods in reconstruction accuracy but also achieves remarkable efficiency gains by significantly reducing computational resource consumption. Operating with approximately 10 GB of GPU memory, NeRIF's compact reconstruction approach enables the spatial resolution of the reconstructed phantom to match that of the captured images closely, facilitating enhanced efficiency. As evidenced in Fig.~\ref{fig:error_time}(b), NeRIF leverages the parallel computing capabilities of GPUs to dramatically accelerate the reconstruction process, reducing the total time from over 10 hours to just over 20 minutes. 

\begin{table}
 \caption{Correlation Coefficients for Various Phantoms and Noise Levels}
  \begin{center}
    \begin{tabular}{lcccccc}
      \hline\hline
      Phantom & Noise 0 & Noise 0.1 & Noise 0.15 & Tikhonov & Landweber \\[1pt]
      \hline
      Phamton1 & 0.9997& 0.9996& 0.9992& 0.9917& 0.9931\\
      Phamton2 & 0.9944& 0.9964& 0.9958& 0.6526& 0.7600\\
      Phamton3 & 0.9981& 0.9969& 0.9971& 0.8259& 0.8809\\
      Phamton4 & 0.9984& 0.9984& 0.9984& 0.8806& 0.9155\\
      Phamton5 & 0.9987& 0.9979& 0.9982& 0.8933& 0.9256 \\
      Phamton6 & 0.9966& 0.9980& 0.9973& 0.9012& 0.9316\\
      Phamton7 & 0.9981& 0.9981& 0.9983& 0.9092& 0.9370\\
      \hline\hline
    \end{tabular}
    
    \label{tab:corr_coeffs}
  \end{center}
\end{table}

\begin{table}
  \begin{center}
    \caption{\(L2\) Errors for Various Phantoms and Noise Levels}
    \label{tab:l2_errors}
    \begin{tabular}{lcccccc}
      \hline\hline
       Phantom & \multicolumn{1}{c}{Noise 0} & \multicolumn{1}{c}{Noise 0.1} & \multicolumn{1}{c}{Noise 0.15} & \multicolumn{1}{c}{Tikhonov} & \multicolumn{1}{c}{Landweber} \\[1pt]
      \hline
      Phamton1 & 0.585\%& 0.817\%& 0.821\%& 2.140\%& 1.820\%\\
      Phamton2 & 0.423\%& 0.419\%& 0.343\%& 2.761\%& 2.360\%\\
      Phamton3 & 0.394\%& 0.503\%& 0.610\%& 3.406\%& 2.857\%\\
      Phamton4 & 0.485\%& 0.590\%& 0.438\%& 3.477\%& 2.948\%\\
      Phamton5 & 0.456\%& 0.542\%& 0.501\%& 3.734\%& 3.143\%\\
      Phamton6 & 0.783\%& 0.639\%& 0.810\%& 4.023\%& 3.372\%\\
      Phamton7 & 0.906\%& 0.837\%& 1.041\%& 4.304\%& 3.607\%\\
      \hline\hline
    \end{tabular}
  \end{center}
\end{table}

\subsection{Experimental  validation}
\begin{figure}
  \centering
  \includegraphics[width=1\linewidth]{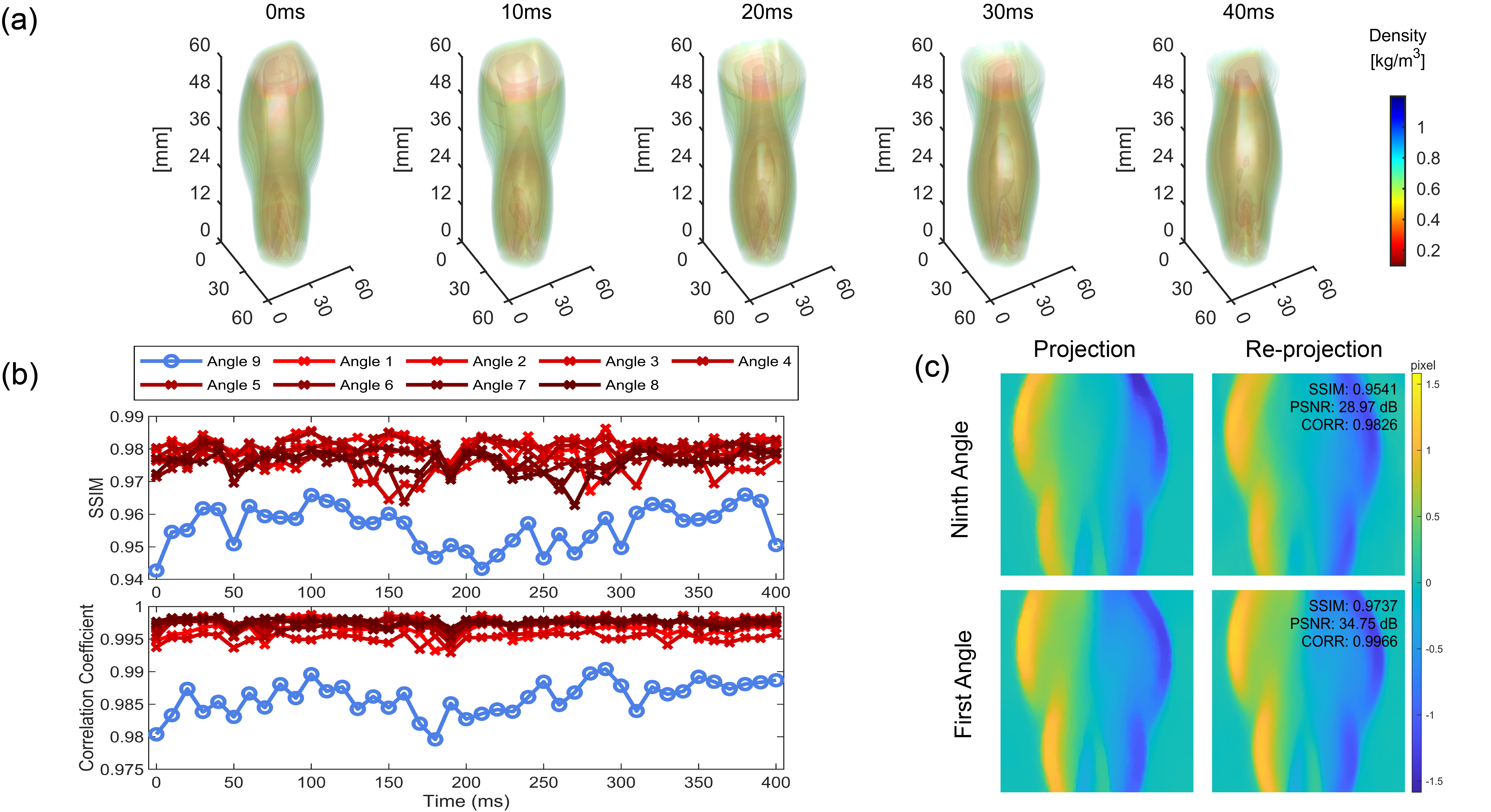}  
  \caption{(a) Reconstruction of the 3D density field experiment from 0 to 40 ms; (b) SSIM and CC of re-projection between the test view and the reconstructed view; (c) Visualization of re-projection errors.}
  \label{fig:re-proj}
\end{figure}

The numerical validation presented above has confirmed the robustness of the NeRIF technique. Now, we provide an experimental demonstration of this approach. Our study focused on a turbulent premixed methane/air flame with an equivalence ratio of $\phi=1.2$. Within the reactor's central nozzle, depicted in Fig.~\ref{fig:setup}, the fuel-air mixture was ignited amidst a nitrogen co-flow. Utilizing the NeRIF technique, we reconstructed the temporal evolution of the three-dimensional density field, as illustrated in Fig.~\ref{fig:re-proj}(a). This reconstruction effectively circumvented the overlap of flame surface structures, enabling \emph{simultaneous} visualization of the flame front leading edge and the interface between the hot combustion products and the nitrogen co-flow.

Given the absence of ground truth data for the direct comparison, we validated our reconstruction with an alternative approach. Out of the nine projections captured, eight were employed for the reconstruction process and one was saved for the validation. Fig.~\ref{fig:re-proj}(b-c) showcase a comparison between the 9th measured projection and the re-projection simulated with the reconstructed refractive index field, in terms of metrics such as SSIM, PSNR, and CC. In order to explain the quality of the reconstructed image, we introduced the PSNR(Peak Signal to Noise Ratio), and detailed information is provided in Appendix.~\ref{app:Metrics}.

\begin{figure}
  \centering
  \includegraphics[width=1\linewidth]{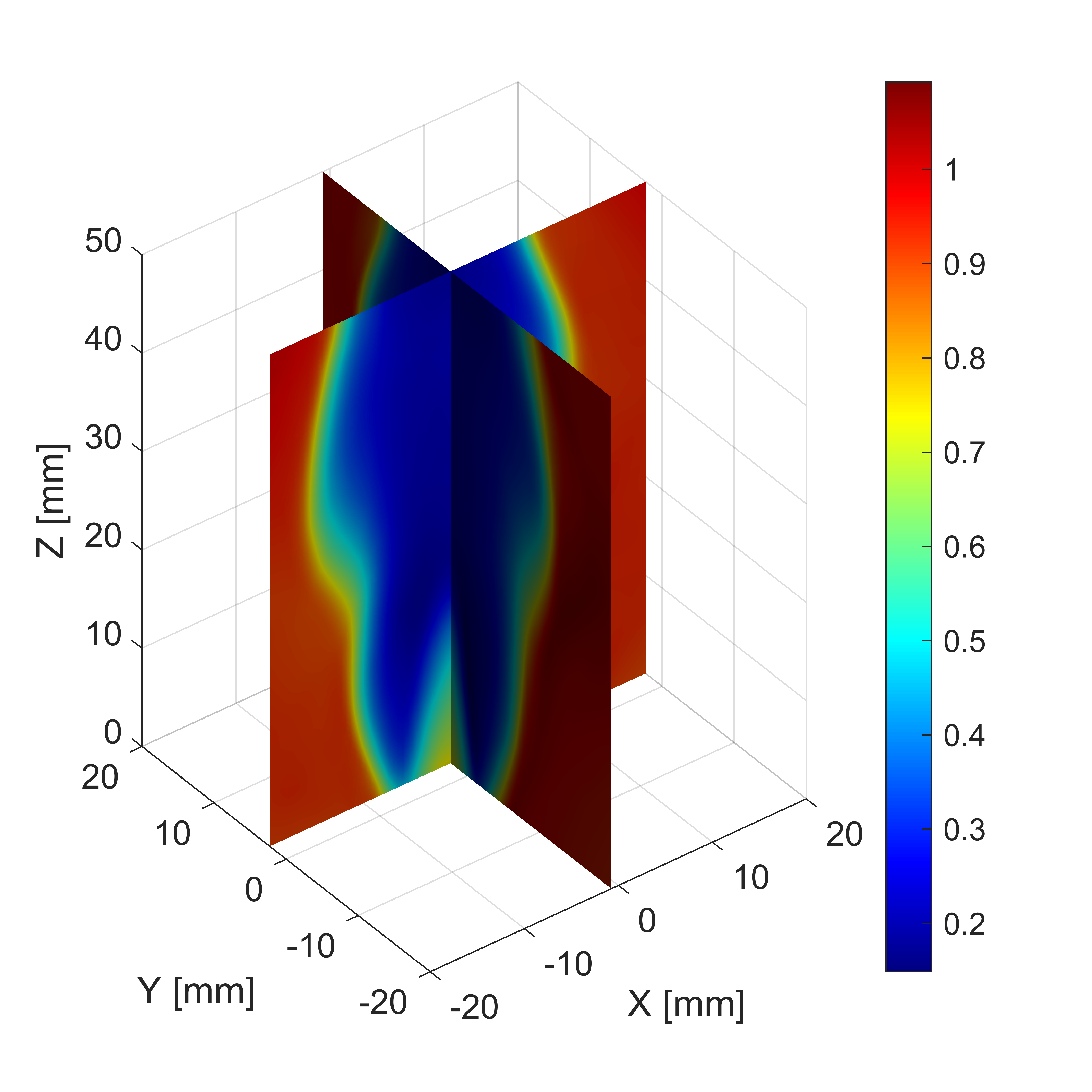}  
  \caption{Visualization of X-axis and Y-axis slices of flame at 0 ms }
  \label{fig:visualxy}
\end{figure}

For eight angles employed in the reconstruction process, the PSNR consistently surpasses 30 when comparing the measured projections with their corresponding re-projections across different phantoms. Furthermore, the SSIM exceeds 0.95, and the CC stands above 0.99, demonstrating a remarkable level of fidelity in the reconstructed three-dimensional density field. Taking the 0ms flame as an example, in Fig.~\ref{fig:visualxy} we visualized the internal structure of the reconstructed flame, and we can find that the unburned low-temperature region of the flame is well reconstructed, and has a clear flame interface. 
We also provide a video demonstration of the reconstructed 3D density field from 0 to 400 ms, available in Movie 1. As a result, the reconstruction can be considered as a highly reliable algorithm and utilized for further investigations on the intricate mechanism of turbulent flames. We also provide videos of Three-dimensional visualization of flame profiles with higher turbulence in Movie 2 and Movie 3.

\section{Conclusions}
In this endeavor, we have successfully demonstrated an innovative approach \emph{i.e.} Neural Refractive Index Field (NeRIF) tailored for volumetric Background-Oriented Schlieren Tomography (VBOST). The NeRIF features several pivotal advantages. Firstly, it achieves a high spatial resolution by leveraging a neural network to represent the flow field as a continuous function, avoiding voxel-based discretization and enabling the spatial resolution to surpass the limitation imposed by the number of voxels and approaching the limit imposed by camera pixel resolution. Consequently, such volumetric representation offers fine-grained control over the flow of intricate structures, allowing the model to capture details with exceptional clarity. Secondly, the random sampling strategy adopted to approximate the continuous integration and differentiation effectively mitigates discretization errors, leading to a significant enhancement in the reconstruction accuracy. Thirdly, NeRIF's execution process eliminates the need for the huge coefficient matrix, which is typically indispensable in voxel-based methods, thereby drastically reducing memory requirements. Lastly, the carefully crafted loss function minimizes inherent mathematical discrepancies between the output refractive index and its gradients, effectively preventing it from over-fitting the measurement noise. By harnessing the flexibility of the neural network fitting function, advancements such as non-linear ray tracing and non-paraxial optical assumptions can be seemingly integrated, further enhancing the reconstruction accuracy. 

In summary, the NeRIF signifies a noteworthy advancement in BOST, offering great prospects for three-dimensional flow and combustion diagnostics applications. More importantly, though demonstrated in the context of BOST, the core mechanisms of NeRIF are readily adaptable to diverse tomographic modalities, including tomographic absorption spectroscopy and tomographic particle imaging velocimetry, thereby extending its potential reach across various domains of fluid visualization and analysis.

\begin{acknowledgments}
This work was funded by National Science Foundation of China under grant number 51706141, 51976122 and SJTU Global Strategic Partnership Fund (2023 SJTU-UoE).
We would like acknowledge Prof. Samuel J. Grauer for the valuable suggestions.
\end{acknowledgments}

\section*{Data Availability Statement}
The data that support the findings of this study are available from the corresponding author upon reasonable request.

\appendix
\section{\label{app:NNstructure}Structure of Neural Network}
The structure of the neural network in this study can be shown in the following table.

\begin{figure}
  \centering
  \includegraphics[width=1\linewidth]{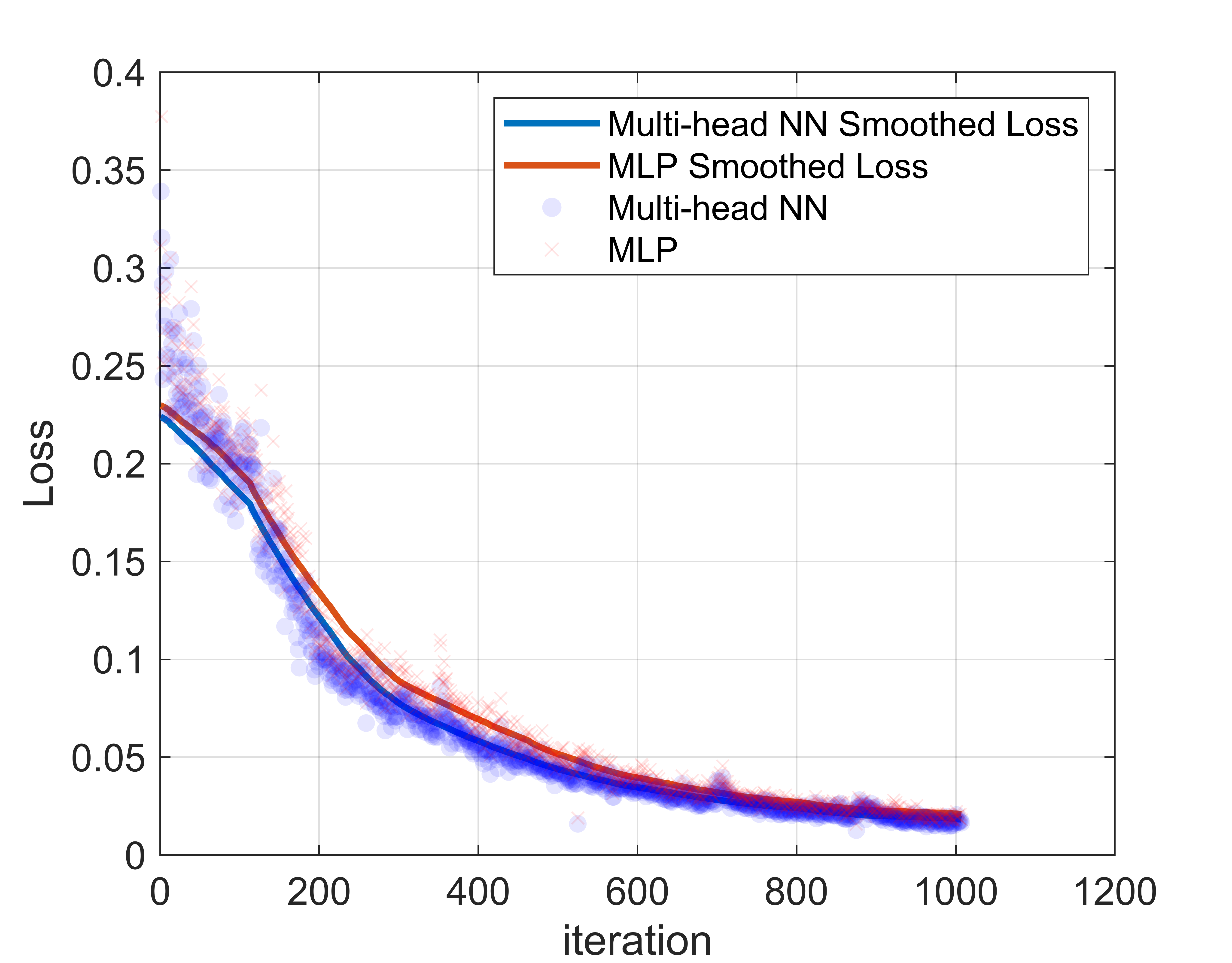}  
  \caption{The influence of neural network structure on reconstruction}
  \label{fig:NNstruct}
\end{figure}

\begin{table}
\centering
\caption{Structure of the Neural Network.}
\begin{tabular}{l|l}
\hline
\textbf{Layers} & \textbf{Dimension} \\
\hline
Input layer    & \(3 \times (2N+1)\) \\
Layer 1        & 256 \\
Layer 2        & 256 \\
Layer 3        & 256 \\
Layer 4        & 256+\(3 \times (2N+1)\)  \\
Layer 5        & 256 \\
Layer 6        & 256 \\
Layer 7        & 256 \\
Layer 8        & 64+64+64+64 \\
Output layer   & 1+1+1+1 \\
\hline
\end{tabular}
\label{tab:NNparameter}
\end{table}
The four outputs are the refractive index \(n\) and its gradient \(\nabla n\) in directions in x, y and z. Here we compare the performance of directly fully connected neural network and our adopted network.In this study, the influence of the last layer structure is tested experimentally, and the convergence characteristics of the Multi-head NN used in this paper are compared with the multi-layer perceptron MLP connected by jumps. As shown in Fig.~\ref{fig:NNstruct}, the structure we adopted has better convergence characteristics, with the mean value of loss being 0.07 and the mean value of MLP loss being 0.8.

\begin{figure}
  \centering
  \includegraphics[width=1\linewidth]{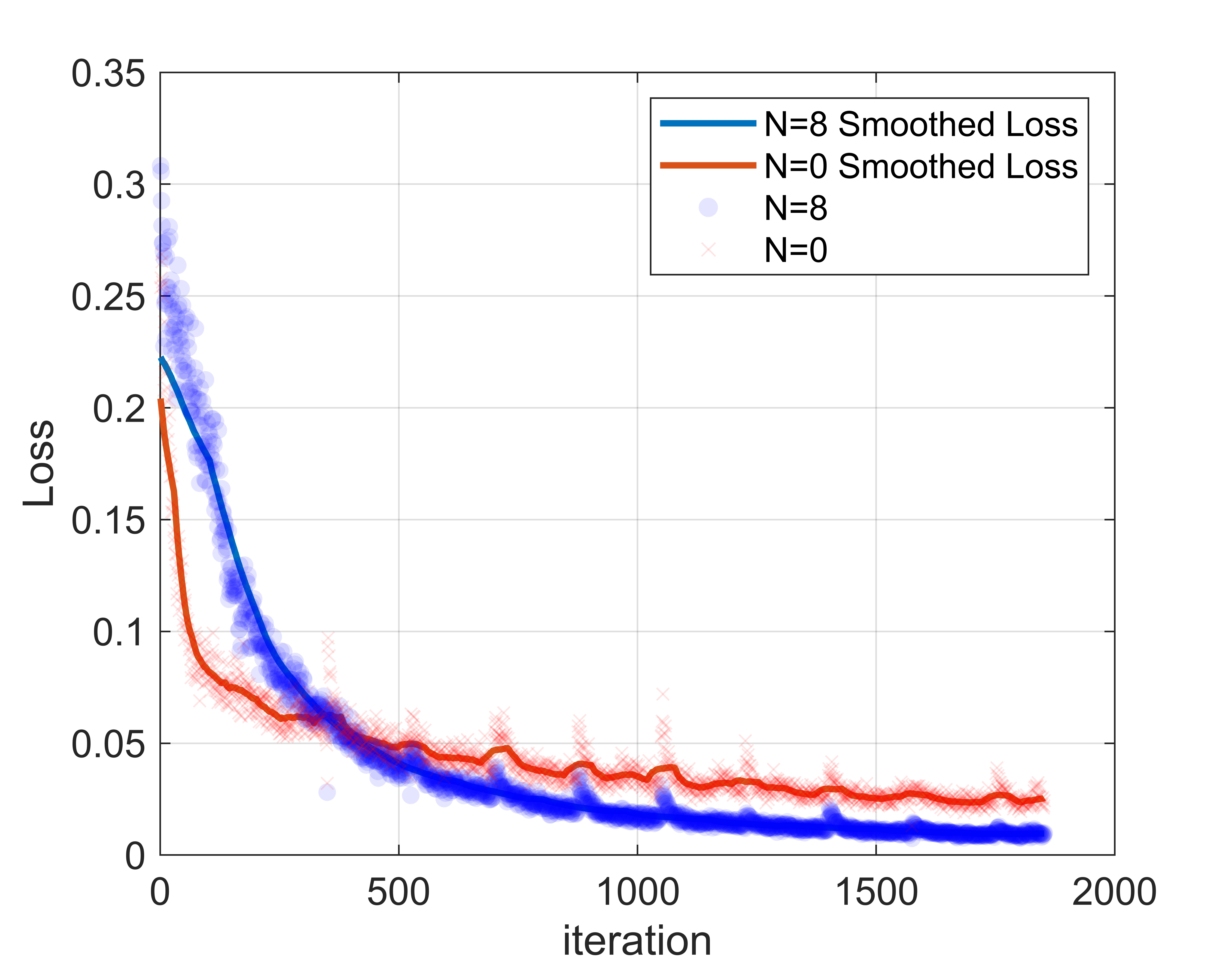}  
  \caption{Effect of position encoding on reconstruction}
  \label{fig:NNencode}
\end{figure}

The positional encoding formula presented in Eq.~\ref{eq:postionencode} is designed to inject relative or absolute position information into the model, where N is the parameter selected according to the complexity of the reconstructed field. Such a coding method can avoid excessive smoothing caused by implicit regularization of the neural network so that the neural network can fit the high-frequency information of the flow field in the early stage, and improve the convergence speed and accuracy. As shown in Fig.~\ref{fig:NNencode}, it can be seen that due to the function of coordinate coding, loss decreases slightly slowly in the initial stage of reconstruction, but the speed of decrease is more uniform and the expected loss threshold can be reached faster in the end, but it is worth noting that excessive coordinate coding may also lead to overfitting of the refractive index field.

\begin{equation}
\label{eq:postionencode}
E(p) = \left( \sin\left(2^0 \pi {p}\right), \cos\left(2^0 \pi {p}\right), \dots, \sin\left(2^{N-1} \pi {p}\right), \cos\left(2^{N-1} \pi {p}\right) \right)
\end{equation}

\section{\label{app:RK}Ray Tracing for Simulation}
\begin{figure}
  \centering
  \includegraphics[width=1\linewidth]{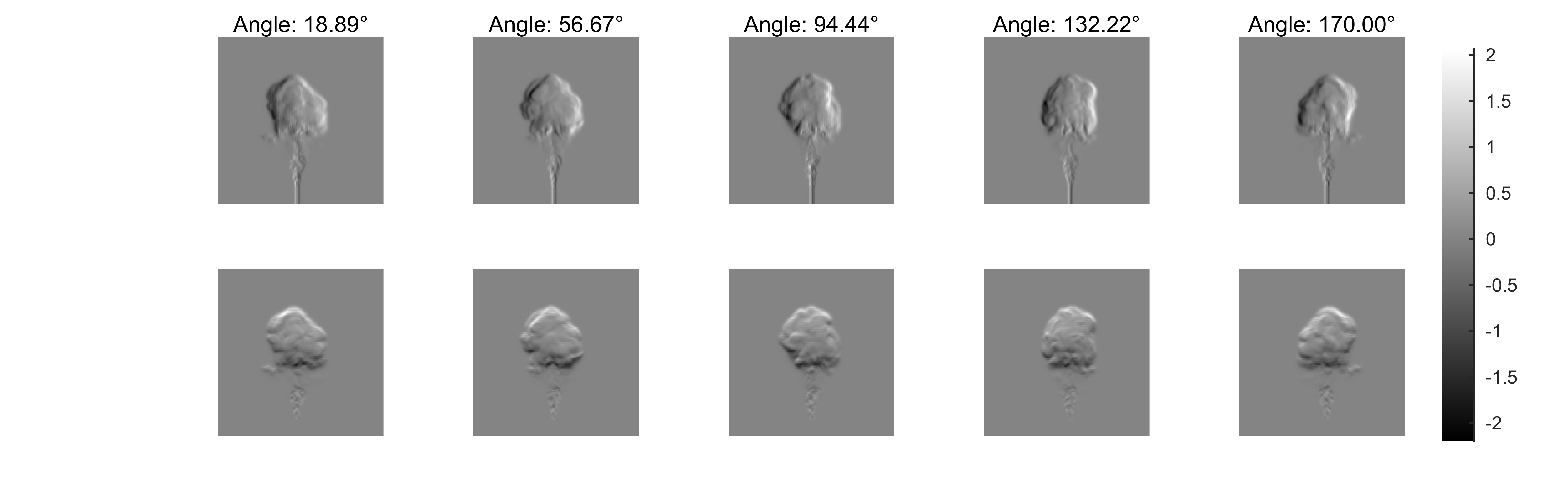}  
  \caption{Projected displacement of spray combustion obtained by simulation}
  \label{fig:deta10ms}
\end{figure}

During the simulation, we used the fourth-order accuracy of the Runge-Kutta method for ray tracing from the camera plane to obtain the propagation of the rays in a non-uniform refractive index field and the displacement of the camera rays.We introduce the position matrix $R$ and the ray vector $T$:\\
\begin{equation}
R = \begin{pmatrix}
x \\
y \\
z
\end{pmatrix}
\end{equation}
\begin{equation}
T = \frac{dr}{dt} = n(r) \frac{dr}{ds} = n(r) \begin{pmatrix}
\cos \alpha \\
\cos \beta \\
\cos \gamma
\end{pmatrix}
\end{equation}
Define the matrix $D$:\\
\begin{equation}
D = n(r) \nabla n(r) = \frac{1}{2} \nabla n(r)^2 = \frac{1}{2} \begin{pmatrix}
\frac{\partial n^2}{\partial x} \\
\frac{\partial n^2}{\partial y} \\
\frac{\partial n^2}{\partial z}
\end{pmatrix}
\end{equation}
Combining equations, we get:
\begin{equation}
T' = D(r)
\end{equation}
Using the Runge-Kutta method, we compute:\\
\begin{equation}
\begin{aligned}
K_1 &= \Delta t D(R_n) \\
K_2 &= \Delta t D\left(R_n + \frac{\Delta t}{2} T_n + \frac{\Delta t}{8} K_1 \right) \\
K_3 &= \Delta t D\left(R_n + \frac{\Delta t}{2} T_n + \frac{\Delta t}{2} K_2 \right) \\
K_4 &= \Delta t D\left(R_n + \frac{\Delta t}{2} T_n + \frac{\Delta t}{2} K_3 \right) \\
R_{n+1} &= R_n + \frac{\Delta t}{6} (6T_n + K_1 + 2K_2 + 2K_3 + K_4) \\
T_{n+1} &= T_n + \frac{\Delta t}{6} (K_1 + 4K_2 + K_3 + K_4)
\end{aligned}
\end{equation}
Here, $\Delta t$ is the step size. The Runge-Kutta method computes the coordinates and direction of the ray iteratively, obtaining multiple coordinate points along the ray's path. This method does not require consideration of the direction and position at each refractive index gradient boundary. Connecting these points easily reveals the ray's trajectory through the gradient index medium. As an example, the displacement obtained by tracing phantom2 is shown in Fig.~\ref{fig:deta10ms}.

\section{\label{app:Metrics}Image Quality Assessment Metrics}
 We use three image quality assessment metrics in this study: Structural Similarity Index Measure (SSIM), Correlation Coefficient (CC), and Peak Signal-to-Noise Ratio (PSNR). These metrics are widely used in the field of image processing to evaluate the quality of reconstructed images.

\subsection{Structural Similarity Index Measure (SSIM)}
The Structural Similarity Index Measure (SSIM) is used for measuring the similarity between two images. It is a perception-based model that considers image degradation as a perceived change in structural information, while incorporating important perceptual phenomena, including both luminance masking and contrast masking terms.

\begin{equation}
\text{SSIM}(x, y) = \frac{(2\mu_x \mu_y + c_1)(2\sigma_{xy} + c_2)}{(\mu_x^2 + \mu_y^2 + c_1)(\sigma_x^2 + \sigma_y^2 + c_2)}
\label{eq:ssim}
\end{equation}

In Eq.~\ref{eq:ssim}, \(\mu_x\) and \(\mu_y\) are the averages of \(x\) and \(y\) respectively, \(\sigma_x^2\) and \(\sigma_y^2\) are the variances, and \(\sigma_{xy}\) is the covariance. The constants \(c_1\) and \(c_2\) are used to stabilize the division with weak denominator.

\subsection{Correlation Coefficient (CC)}
The Correlation Coefficient is a statistical measure that calculates the strength of the linear relationship between two variables. It is dimensionless and varies between -1 and +1, where 1 means a perfect positive linear relationship, -1 means a perfect negative linear relationship, and 0 means no linear relationship at all.

\begin{equation}
\text{CC}(X, Y) = \frac{\sum (X_i - \bar{X})(Y_i - \bar{Y})}{\sqrt{\sum (X_i - \bar{X})^2 \sum (Y_i - \bar{Y})^2}}
\label{eq:cc}
\end{equation}

\subsection{Peak Signal-to-Noise Ratio (PSNR)}
PSNR is most commonly used to measure the quality of reconstruction of lossy compression codecs (e.g., for image compression). The signal in this case is the original data, and the noise is the error introduced by compression.

\begin{equation}
\text{PSNR} = 10 \cdot \log_{10} \left(\frac{\text{MAX}_I^2}{\text{MSE}}\right)
\label{eq:psnr}
\end{equation}
Here, \(\text{MAX}_I\) is the maximum possible pixel value of the image, and MSE stands for Mean Squared Error between the original and compressed image.

\section{\label{app:spatialreslution}Spatial Resolution Determination}
\begin{figure}
  \centering
  \includegraphics[width=1\linewidth]{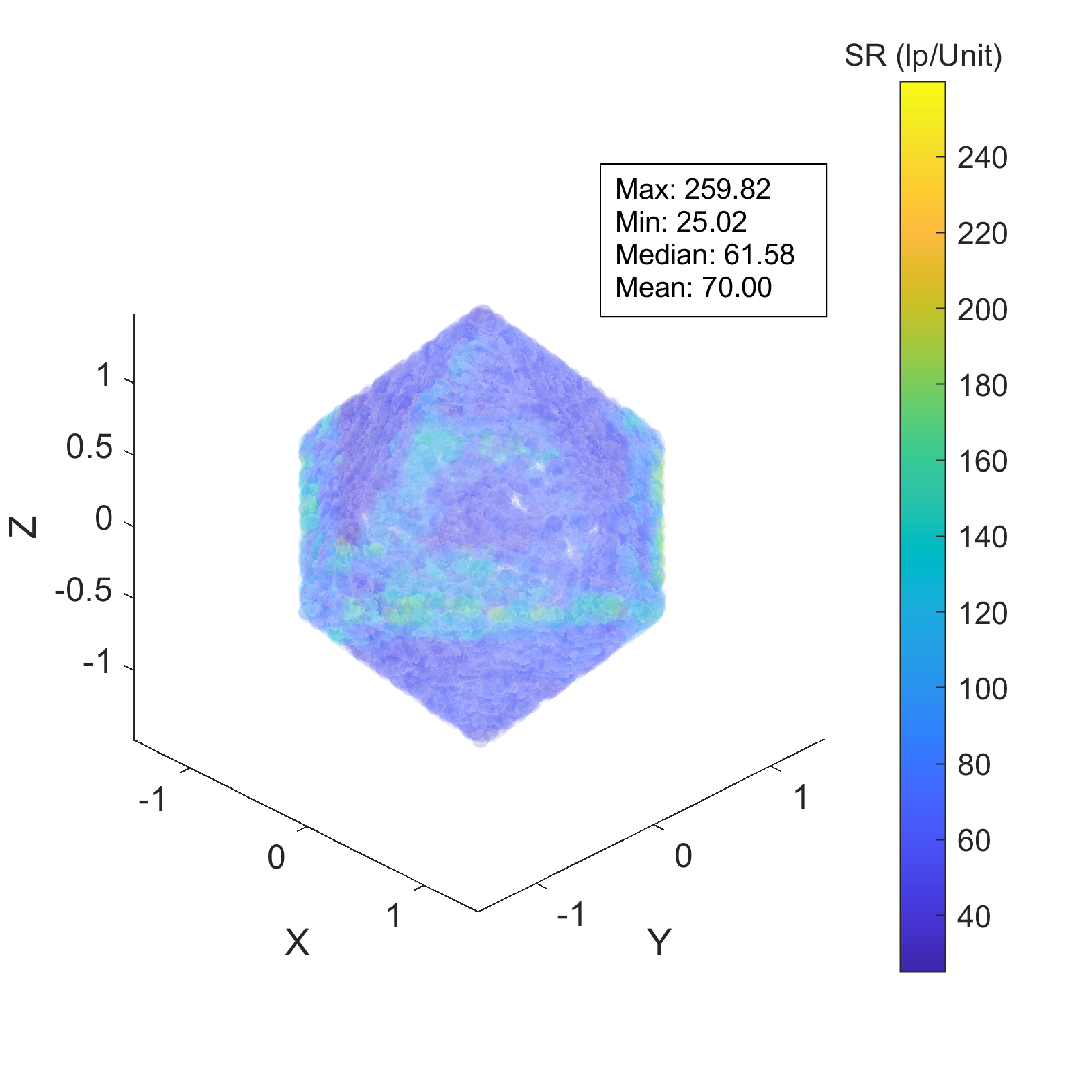}  
  \caption{Distribution of three-dimensional spatial resolution}
  \label{fig:SR}
\end{figure}

There are various ways to define spatial resolution (SR). In this work, we adopt a definition based on the modulation transfer function (MTF) \citep{yu2017quantification}. The line spread function (LSF) is calculated by taking the first derivative of the edge spread function (ESF). The MTF is then obtained by performing a 1D discrete Fourier transform of the LSF. 
In the numerical experiments, we constructed a piecewise function within a 4 \(\times\) 4 \(\times\) 4 unit space, as shown in Eq.~\ref{eq:SRphantom}. The derivative of this function features sharp boundaries, where the density gradient at the boundary surface is a step function, which can be used for MTF calculation. A 10\% cutoff threshold was applied to the MTF, resulting in an SR of 61.58 lp/unit,The distribution of 3D spatial resolution is shown in Fig.~\ref{fig:SR}.

\begin{equation}
\label{eq:SRphantom}
n(x, y, z) = \begin{cases} 
1.5 - \left(|x| + |y| + |z|\right), & \text{if } |x| + |y| + |z| \leq 1.5 \\
0, & \text{otherwise}
\end{cases}
\end{equation}

\nocite{*}
\bibliography{main}

\end{document}